\documentclass[preprint,%
tightenlines,%
superscriptaddress,%
floats,prd,eqsecnum,nobibnotes%
,nofootinbib,aps,12pt]{revtex4}
\usepackage{amsmath, amssymb}

\newif\ifeprint
\eprinttrue

\newcommand{\beq}{\begin{equation}}
\newcommand{\eeq}{\end{equation}}
\newcommand{\beqa}{\begin{eqnarray}}
\newcommand{\eeqa}{\end{eqnarray}}
\newcommand{\be}{\begin{equation}}
\newcommand{\ee}{\end{equation}}
\newcommand{\bea}{\begin{eqnarray}}
\newcommand{\eea}{\end{eqnarray}}
\newcommand{\bA}{\begin{array}}
\newcommand{\eA}{\end{array}}
\newcommand{\bc}{\begin{center}}
\newcommand{\ec}{\end{center}}
\newcommand{\al}{\alpha}

\newcommand{\ra}{\rightarrow}
\newcommand{\del}{\partial}
\newcommand{\p}{\partial}
\newcommand{\ie}{{\it i.e.}}
\newcommand{\eg}{{\it e.g.}}

\newcommand{\Nf}{${\mathcal N}{=}4$}

\begin{document}

\ifeprint
\setlength{\baselineskip}{1.2\baselineskip} 
\fi


\title{\vspace*{\fill}\LARGE Time Dependent Cosmologies and Their Duals}

\author{Sumit R. Das}\email{das@pa,uky,edu}
\author{Jeremy Michelson}\email{jeremy@pa,uky,edu}
\affiliation{Department of Physics and Astronomy, 
      University of Kentucky, 
      Lexington, KY \ 40506 \ U.S.A.}
\author{K. Narayan}\email{narayan@theory,tifr,res,in}
\author{Sandip P. Trivedi} \email{sandip@theory,tifr,res,in}
\affiliation{Department of Theoretical Physics, 
Tata Institute of Fundamental Research, 
Homi Bhabha Road, Colaba, 
Mumbai - 400005, INDIA.
\vspace*{\fill}} 


\begin{abstract}
\vspace*{\baselineskip}
We construct a family of solutions in IIB supergravity theory. These are time dependent or depend on a  light-like
 coordinate and can be thought of as deformations of $AdS_5\times S^5$. Several of the solutions have 
singularities.  The light-like solutions preserve $8$ supersymmetries. 
We argue that these solutions are dual to the ${\mathcal N}=4$ gauge theory  in a $3+1$ dimensional spacetime  with 
a metric and a gauge coupling that is varying with time or the  light-like direction respectively.
This identification allows us to map the question of singularity resolution to the dual gauge theory.
\vspace*{\fill}
\end{abstract}

\preprint{\parbox[t]{10em}{\begin{flushright}
TIFR-TH-06-05 \\ 
UK/06-03 \\ {\tt hep-th/0602107} 
\end{flushright}}}

\maketitle

\ifeprint
\tableofcontents
\fi


\section{Introduction}

Time dependent phenomena are poorly understood in string theory.  It
is important to understand them better.  This could lead to an improved  
 understanding of the big-bang and black-hole
singularities and  a  better connection between string theory and
observational cosmology. It could also  reveal  how time
originates from a more fundamental description. Earlier work on time
dependent phenomena includes the large body of results on stringy
backgrounds unstable to tachyon condensation. Some previous attempts
at studying time dependent phenomena pertaining to cosmological
singularities in string theory include \eg\ \cite{horowitzsteif, bhkn,
cornalbacosta, lms1, lms2, ckr, keshav0302, grs, bp0307, bdpr,
karczstrom, daskarcz0412, dp,
das0503, mcgreevysilver0506, csv0506, li0506, berkooz0507, lisong0507,
Hikida:2005ec, dasmichel0508, Chen:2005mg, 
she0509, Chen:2005bk, Ishino:2005, robbinssethi0509, KalyanaRama:2005uw, 
hikidatai0510, Li:2005ai, Craps:2006xq, dmichelson2}.

In this paper, in part inspired by \cite{csv0506}, we take a modest
step in trying to understand some time dependent backgrounds in string
theory. We find a family of time dependent and null
backgrounds\footnote{We will loosely call backgrounds which depend on
a light-like coordinate (instead of a time-like coordinate) as null
backgrounds in this paper.}  in Type IIB string theory.  These
solutions are deformations of the $AdS_5\times S^5$ background.
The solutions have singularities which are space-like or null.
In several cases the dilaton is weakly coupled at the singularity.   

We argue that these backgrounds have a dual interpretation in terms of
turning on sources in the ${\mathcal N}=4$ gauge theory. The sources
are time dependent or dependent on the light-like coordinate
respectively.  This dual interpretation allows us to map the question
of the resolution of the singularity to the gauge theory. If the gauge
theory is non-singular in some cases, it should provide an answer to
how the singularity is resolved.

We have not been able to settle this important question in this paper
and postpone a more detailed analysis of it for the future \cite{WiP}.
It is worth mentioning that the null backgrounds are especially
interesting in this context.  These solutions preserve $8$ of the $16$
supersymmetries left unbroken by a D3 brane.  In some cases the
supergravity background corresponds to turning on sources in the gauge
theory which become asymptotically constant, as $X^+ \rightarrow
\pm \infty$. The supergravity solution also corresponds in the gauge
theory to starting, as $X^+ \rightarrow -\infty$, in a state which is
the ${\mathcal N}=4$ vacuum.  In these cases it is possible that a
careful analysis shows that the gauge theory is non-singular and in a
well defined state in the far future, as $X^+ \rightarrow \infty$, as
well.

The  solutions we find are also of interest from the point of view of determining   the 
 response of the  ${\mathcal N}=4$ gauge theory to    time dependent sources.
For this purpose even bulk singular solutions which cannot be resolved might be interesting.
Such singular solutions are  dual to  turning on sources in the gauge theory 
 which became singular at some moment of time. Prior to this moment  it is still valid to 
ask about the response of the gauge theory to the source and this information is contained in the 
supergravity solution.

Using the ideas  of this paper similar solutions can also be obtained in other $AdS$ spaces. 
Particularly interesting is the $AdS_3\times S^3$ case. Here it might be possible to analyse 
some backgrounds, which have a  singularity with a weakly coupled  dilaton, 
using the world sheet conformal field theory.

While this paper was being written,~\cite{chuho0602} appeared.
It contains substantial overlap with the results
presented here.

\section{Supergravity solutions with cosmological singularities}
\label{sec:soln}

We will consider Type IIB supergravity and work in 10-dimensional
Einstein frame.
We are interested in solutions in which the metric, five form, $F_5$, and dilaton, $\phi$, 
are excited. 
Our main result is that any 
background with metric and 5-form 
\begin{equation}
\begin{aligned}
ds^2 &= Z^{-1/2}(x) \tilde{g}_{\mu\nu} dx^\mu dx^\nu
+ Z^{1/2}(x) \tilde{g}_{mn} dx^m dx^n, \\
F_{(5)} &= -\frac{1}{4 \cdot 4!} \tilde{\epsilon}_{\mu\nu\rho\sigma}
  \frac{\p_m Z(x)}{Z(x)^2} dx^\mu \wedge dx^\nu \wedge dx^\rho \wedge dx^\sigma
  \wedge dx^m \\ & \qquad
+ \frac{1}{4\cdot 5!} \tilde{\epsilon}_{m_1m_2m_3m_4m_5}{^{m_6}} \p_{m_6} Z(x)
    dx^{m_1} \wedge dx^{m_2} \wedge dx^{m_3} \wedge dx^{m_4} \wedge dx^{m_5},
\end{aligned}
\end{equation}
is a solution of the equations of motion, as long as
$Z(x)$ is a {\em harmonic\/} function on the flat, six dimensional
tranverse
space with coordinates $x^m$;
$\tilde{g}_{mn}$ is Ricci-flat and depends only on the $x^m$; and
$\tilde{g}_{\mu\nu}$ 
and the dilaton $\phi$ are only dependent on the 4-coordinates $x^\mu$, 
and satisfy the conditions,
\begin{subequations}\label{conds}
\begin{align}
{\tilde R}_{\mu\nu} &= {1\over 2} \partial_\mu\phi \partial_\nu\phi, 
\\
 \partial_\mu (\sqrt{-\det({\tilde g})}\ {\tilde g}^{\mu \nu} 
\partial_\nu \phi) &= 0. 
\end{align}
\end{subequations}%
Upon taking the near horizon limit with
a flat transverse space, this reduces to
\bea
\label{geom}
ds^2=(\frac{r^2}{R^2}){\tilde g}_{\mu\nu} dx^\mu dx^\nu + 
(\frac{R^2}{r^2})dr^2 + R^2 d\Omega_5^2\ , \qquad\
F_{(5)}=R^4(\omega_5 + *_{10} \omega_5)\ ,
\eea
which is the case we will typically consider.
Here it is important to emphasize that ${\tilde R}_{\mu\nu}$ is the Ricci 
curvature made from the metric ${\tilde g}_{\mu\nu}$ alone (without the 
${r^2\over R^2}$ warp factor in front). In Eq.~(\ref{geom}) $d\Omega_5^2$ is the volume element and $\omega_5$ 
is the volume form of the unit five-sphere.
In other words, as long as ${\tilde g}_{\mu\nu}, \phi$, together solve the 
equations of 4-dimensional Einstein gravity coupled to a free scalar field 
in (\ref{conds}), we have a solution to the original 10-dimensional 
problem.

Let us see how this result is obtained. It is easy to see that self-duality 
for the 5-form means that\ \ $F^2\equiv F_{ABCDE}F^{ABCDE}=0$, \ 
where $A, B \ldots$\ take values in ten dimensions. The Einstein equations 
then take the form
\beq\label{eeom}
R_{AB}=\frac{1}{6}
 F_{A A_1A_2A_3A_4}F_B{^{A_1A_2A_3A_4}} + \frac{1}{2}  \partial_A \phi 
\partial_B \phi\ .
\eeq
Now for the background (\ref{geom}) it is clear that this equation with 
components along the $S^5$ directions is satisfied, since the dilaton does not 
depend on the angular coordinates of the $S^5$. The problem then  reduces 
to studying this equation with  components along the $t,x^1,\cdots x^3,r$ 
directions. 

The 5-dimensional metric along these directions can be written as follows:
\beq\label{fdimmet}
ds^2={R^2\over z^2} ({\tilde g}_{\mu \nu}dx^\mu dx^\nu + dz^2)
\eeq 
where $z=R^2/r$. Now using the standard rules 
relating the  Ricci curvature tensor for two metrics related by a conformal 
transformation (see \eg\ \cite{wald}, Appendix D) we then get that 
\be\label{compcur}
R_{\mu\nu} = {\tilde R}_{\mu \nu} -{4 \over R^2} g_{\mu \nu}\ , \qquad 
R_{zz} = -{4 \over R^2} g_{zz}\ .
\ee
It is easy to see that the term $-{4 \over R^2} g_{\mu\nu}$ in the first equation is 
canceled by the 5-form contribution (which in effects provides a 
negative cosmological constant in 5-dimensions). 
Thus we see that as long as ${\tilde R}_{\mu\nu}$ meets the condition in 
(\ref{conds}), the Einstein equations with components along $\mu,\nu$ 
directions are satisfied. 
The Einstein equation component along the $zz$ directions is met because 
the $R_{zz}$ contribution is again balanced by the 5-form flux. Finally, 
the dilaton equation in (\ref{conds}) then follows by noting that it 
satisfies the massless free-field equation in 10 dimensions  and is independent of $z$. 

Let us end this subsection by noting that one can obtain
$AdS_5 \times S^5$ as a special case of the solutions Eqs.~\eqref{geom}
and~\eqref{conds} by taking
$\phi$ to be constant and ${\tilde g}_{\mu\nu}=\eta_{\mu\nu}$.
There are also linearised small fluctuations about this solution which are included in  Eqs.~\eqref{geom} and \eqref{conds}.
Perturbations of the form, $\phi =\phi_0 + \delta \phi(x^\mu)$, ${\tilde g}_{\mu\nu}=\eta_{\mu\nu}$, where
$\phi_0$ is constant and $\delta \phi(x^\mu)$ satisfies Eq.~(\ref{conds}), is of this type.
Similarly included are metric perturbations of the form, ${\tilde g}_{\mu\nu}=\eta_{\mu\nu} + h_{\mu\nu}$, with constant 
dilaton and vanishing linearised Ricci tensor ${\tilde R}_{\mu\nu}$.

\subsection{Time-dependent cosmological singularities}

We now turn to exploring various solutions. In this subsection we
construct time-dependent solutions.

As an example consider  the spacetime, 
\be\label{ega}
{\tilde g}_{\mu\nu}dx^\mu dx^\nu=-dt^2+\sum_{i=1}^3 t^{(2p_i)} dx^idx^i.
\ee
It has the non-zero Ricci components
${\tilde R}_{tt}=(\sum_i p_i -\sum_ip_i^2)/t^2\ , \ 
{\tilde R}_{ii}=p_i(\sum_j p_j-1)t^{2p_i}/t^2$.
The conditions, (\ref{conds}) are solved if 
\be\label{egconds}
\phi =  \alpha \log t\ , \qquad \sum_i p_i = 1\ , \qquad 
{\alpha^2 \over 2} = 1 - \sum_i p_i^2\ .
\ee
One simple set of examples are the Kasner geometries where the dilaton is 
constant and $\sum_i p_i=\sum_i p_i^2=1$. Another more symmetric example 
is $p_i={1\over 3}$ and $\alpha={2\over \sqrt{3}}$ . Note that in this 
case the metric ${\tilde g}_{\mu \nu}$ can be made conformally flat after 
a redefinition of the time coordinate. Starting with a metric of the form,
\be
d{\tilde s}^2 = e^{f(T)}
(-2dT^2 + e^{H_1(T)}dx_1^2 + e^{H_2(T)}dx_2^2 + e^{H_3(T)}dx_3^2)\ ,
\ee
one can show that the metric, Eq.~(\ref{ega}), meeting conditions 
Eq.~(\ref{egconds}), is the most general solution for the conditions, 
Eq.~(\ref{conds}). 

Note that the solutions,  Eqs.~\eqref{ega} and~\eqref{egconds},
have a curvature singularity at $t =0$.\footnote{
In fact in this case both the Einstein frame and string frame metrics have 
a singularity  where the Ricci scalar blows up at $t=0$. }
The string coupling is given by $g_s=e^\phi$. Depending on the sign chosen 
for the dilaton in Eq.~\eqref{egconds} the string coupling blows up or goes 
to zero at the singularity. It is easy to verify that the curvature 
singularity is reached in finite proper time for a time-like geodesic.

The class of time-dependent solutions allowed by Eq.~(\ref{conds}) is
in fact much larger.  With constant dilaton, ${\tilde g}_{\mu\nu}$ can
be the metric of a gravitational wave, or that of the Schwarzschild
black hole. In the latter case the region inside the horizon gives a
time-dependent geometry with a big-bang or big-crunch.  Other
solutions include ${\tilde g}_{\mu \nu}$ being a homogeneous FRW
metric sourced by an appropriate time-dependent dilaton. These again
have big-bang and big-crunch singularities.

With Euclidean signature again several solutions are possible. For
example, with constant dilaton, any Ricci flat metric, ${\tilde
g}_{\mu \nu}$, is allowed.


We started with an $S^5$ in Eq.~(\ref{geom}), but analogous solutions
can be obtained by replacing it with an constant curvature compact
five manifold. An example is the base of the conifold, $T^{1,1}$.
Additional solutions can also be obtained by using duality. For
example using the S-duality of the IIB theory, one can obtain
solutions where the axion is also turned on.\footnote{We mention here 
the papers \cite{keshav0302, keshav0501, Ishino:2005} that find 
families of supergravity solutions in M-theory with dependences on 
either timelike or lightlike time coordinates: it would be interesting 
to explore the relations between these and our solutions here in the 
Type IIB context, and generalize them.}

The time-dependent solutions in this section of course do not preserve
any supersymmetries.

\subsection{Null cosmological singularities}

Next we turn to null solutions. These depend on one light-like
coordinate which we call $X^+$.

The spacetime takes the form\footnote{There is some redundancy in this
choice of metric components, but we keep it because it will be useful
in the subsequent discussion.}
\bea\label{gennullsoln1}
d{\tilde s}^2 &=& e^{f(X^+)} (-2dX^+ dX^- + e^{H(X^+)} (dX^+)^2 
+ e^{h_2(X^+)} dx_2^2 + e^{h_3(X^+)} dx_3^2)\ , \nonumber\\
{} \phi &=& \phi(X^+).
\eea
The only non-vanishing component of the Ricci tensor is ${\tilde R}_{++}$. 
We should also note that ${\tilde R}_{++}$ is independent of $H(X^+)$.  
The background, Eq.~(\ref{gennullsoln1}), is a solution if it meets the 
condition,
\be\label{gennullsoln2}
{1\over 2}(\del_+ \phi)^2 = {\tilde R}_{++}={1\over 2}(f')^2 - f'' - 
{1\over 4} [(h_1')^2 + (h_2')^2] - {1\over 2} (h_1'' + h_2'')\ , 
\ee
where\ $h_1'={dh_1\over dX^+}$ etc.  
This is a large three function-parameter family of solutions. By a  
coordinate transformation we can set  $f(X^+)=1$.
For any choice of $H(X^+), h_2(X^+), h_3(X^+)$ we can then get a solution 
by choosing a dilaton which satisfies Eq.~(\ref{gennullsoln2}).

For simplicity, let us focus on spacetimes conformal to flat space: thus 
we set $g_{++}=e^H=0$ and $h_2=h_3=0$ and the background becomes
\be\label{nullflat}
d{\tilde s}^2 = e^{f(X^+)} (-2dX^+ dX^- + dx_2^2 + dx_3^2)\ , \qquad
\phi=\phi(X^+)\ .
\ee
This is a solution if 
\be
\label{condnull}
{1\over 2}(f')^2 - f'' = {1\over 2}(\del_+ \phi)^2\ .
\ee
The 10D Einstein frame background 
(\ref{geom}) has the curvatures 
\be
R=0\ , \qquad R_{++}={1\over 2} (f')^2 - f''\ , \qquad 
-R_{+-}=R_{22}=R_{33}= {4 e^{f(X^+)} r^2\over R^4}\ ,
\ee
besides the ones with components on $S^5$. 
Note that if $e^f \rightarrow 0$ the metric components,  
Eq.~(\ref{nullflat}), shrink to zero.  The conditions for the existence 
of a singularity in such a situation are made  more precise below. 

Null geodesics in the spacetime (\ref{nullflat}) at constant $X^-,x^2,x^3$,
\ie\ trajectories moving only along $X^+$, satisfy the condition,
\be
{d^2X^+\over d\lambda^2} + \Gamma^+_{\al\beta} {dx^{\al}\over d\lambda}
{dx^{\beta}\over d\lambda} =
{d^2X^+\over d\lambda^2} + f' \biggl({dX^+\over d\lambda}\biggr)^2
= 0\ ,
\ee
since the only nonzero $\Gamma^+_{\al\beta}$ is $\Gamma^+_{++}=f'$.
Here $\lambda$ is the affine parameter for the geodesic. 
This can be solved to give $\lambda$ in terms of $X^+$,
\be
\label{deflambda}
\lambda = {\rm const.}\ \int e^{f(X^+)} dX^+\ .
\ee
The curvature component, 
\be
\label{affcurv}
R_{\lambda \lambda}=R_{++} \left({dX^+ \over d\lambda}\right)^2 = 
\left({1\over 2} (f')^2-f^{''}\right) e^{-2f}\ .
\ee
For a suitably chosen $f$ this can blow up when $e^f \rightarrow 0$. 
As we will see in the examples below this can occur at
a finite value of $\lambda$. 

Our  first example is obtained by taking  $f(X^+)=-QX^+$. Then the dilaton 
takes the form, $\phi=\pm QX^+$. We find from Eq.~(\ref{affcurv}) that 
$R_{\lambda \lambda}={1\over 2} Q^2 e^{2QX^+}$. This blows up at
$X^+ \rightarrow \infty$, showing that there is a big crunch curvature 
singularity. The singularity occurs at finite affine parameter,
 $\lambda = e^{-QX^+} \rightarrow 0$, as $X^+\rightarrow \infty$.  
This spacetime, with a curvature  singularity at finite affine parameter, 
is thus geodesically incomplete.
Choosing the dilaton to be $\phi=-QX^+$ we find that the string coupling 
constant vanishes at the singularity.%
\footnote{$R_{\lambda \lambda}$ in the string frame also blow up at the singularity in this case.}

Our next example is obtained by taking $e^{f(X^+)}=\tanh^2X^+$. 
This gives, 
\be\label{nullsolntanh}
d{\tilde s}^2 = \tanh^2 X^+ (-2dX^+ dX^- + dx_2^2 + dx_3^2)\ , 
\qquad e^{\phi}=g_s \biggl(\tanh {X^+\over 2}\biggr)^{\sqrt{8}}\ ,
\ee
The example has been  engineered so that the spacetime becomes flat in the far
past and future (with constant dilaton) but exhibits interesting
behaviour in the intermediate region. We have 
$R_{++}={4\over\sinh^2 X^+}$, so that
$R_{\lambda\lambda}=\frac{4}{\sinh^2X^+ \tanh^4X^+}$
 showing a curvature singularity at 
$X^+\ra 0$, with $e^{\phi}$ becoming arbitrarily small there.%
\footnote{The string frame curvature, $R_{\lambda \lambda}$ blows up at 
the singularity in this case as well.}
It is easy to see from Eq.~(\ref{deflambda}) that the singularity occurs
at finite value of the affine parameter.  

We end with two comments.  
First, it is worth pointing out that the only solutions to 
(\ref{gennullsoln2}) with a constant  constant dilaton is flat space. 
With $\phi=\text{const}$, (\ref{gennullsoln2}) leads to 
$e^f={1\over (X^+)^2}$. After a coordinate transformation this can be put 
in the form of the standard flat space metric.

Secondly, introduction of a nontrivial ${\tilde g}_{\mu\nu}(x^\nu)$ in 
(\ref{geom}) typically introduces curvature singularities at 
the Poincare horizon at $r=0$. However it turns out for the null
solutions considered above there is no such singularity. This is yet
another reason why we focus on the null solutions.

\subsubsection{Supersymmetry of the null solutions}

In this subsection we explore the supersymmetry of the null solutions. 
For simplicity, we restrict ourselves 
to the solutions, Eqs.~(\ref{geom}) and~(\ref{nullflat}). 

We are considering a Type IIB background with Einstein metric $g_{MN}$, 
dilaton $\Phi$ and 5-form $F_{MNPQR}$, using the notation of 
\cite{granapolch} and the earlier \cite{schwarz}. Since 
$\tau=ie^{-\phi}$, the quantity $B={1+i\tau\over 1-i\tau}$ is real and 
the quantity $Q_M=(1-BB^*)^{-1}{\rm Im}(B\del_M B^*)=0$. Then the 
supersymmetry variations are
\begin{subequations}
\begin{gather}
\delta\lambda={i\over\kappa}\gamma^M (1-B^2)^{-1}\del_MB\varepsilon^* 
= 0, \\
\delta\psi_M={1\over\kappa}D_M\varepsilon + \frac{i}{480}
\gamma^{M_1\ldots M_5}F_{M_1\ldots M_5}\gamma_M\varepsilon = 0,
\end{gather}
\end{subequations}%
for the complex Weyl dilatino $\lambda$ ($\gamma^{11}\lambda=\lambda$) 
and the complex Weyl gravitino $\psi_M$ ($\gamma^{11}\psi=-\psi$).
Also the supersymmetry parameter $\varepsilon$ is a complex Weyl spinor 
with $\gamma^{11}\varepsilon=-\varepsilon$. The covariant derivative is\ 
$D_M=\del_M + \frac{1}{4}\omega^{ab}_M \Gamma_{ab}$, with 
$\Gamma_{ab}=\frac{1}{2}[\Gamma_a,\Gamma_b]$,
using flat space $\Gamma$-matrices and curved space $\gamma$-matrices
which satisfy $\gamma_M=e^a_M\Gamma_a$.
The 
spin connection $\omega^{ab}$ satisfies\ 
$de^a + \omega^a_{\ b}\wedge e^b=0$, the $e^a$ being an orthonormal frame.

The null background, Eqs.~(\ref{geom}) and~(\ref{nullflat}), take the form,
\begin{subequations}\label{gensolnZ}
\begin{gather}
ds_E^2 = Z^{-1/2}e^f (-2dX^+dX^-+dx^idx^i)+Z^{1/2}dx^mdx^m, \\
F_{0123m} = \frac{1}{4\kappa} \frac{1}{Z e^{-2f}} \del_m\log Z, 
\qquad F_{m_1m_2m_3m_4m_5} = \frac{1}{4}
              \epsilon_{m_1m_2m_3m_4m_5}{^{m_6}} \p_{m_6} Z,
\qquad \phi=\phi(X^+),
\end{gather}
\end{subequations}%
where $Z=Z(x^m)$ is a harmonic function in the flat transverse space 
with coordinates $x^m$, \ $f=f(X^+)$, and the $\epsilon_{(6)}$ is the 
flat one. In our notation the indices $i=1,2$, refer to two directions 
parallel to the D3 brane, and the indices $m=1, \cdots, 6$ refer to the 
six directions transverse to the D3 brane.
We choose the obvious, diagonal frame $e^+ = Z^{-1/4} e^{f/2} dx^+$, etc.

The spin connection in this background, with the above choice of frame, is, 
\begin{equation} \label{spincon}
\begin{aligned}
\omega_{-+} &= -\frac{1}{2}f' dX^+, &
\omega_{-m} &= \frac{1}{4} Z^{-1/2} e^{f/2} \del_m\log Z dX^+, &
\omega_{mn} &= \frac{1}{4}(\del_n\log Z dx^m - \del_m\log Z dx^n), \\
\omega_{i+} &=\frac{1}{2}f' dx^i, &
\omega_{+m} &= \frac{1}{4} Z^{-1/2} e^{f/2} \del_m\log Z dX^-, &
\omega_{im} &=-\frac{1}{4}Z^{-1/2}e^{f/2}\del_m\log Z dx^i.
\end{aligned}
\end{equation}

It is easy to see that for the background~(\ref{geom}), (\ref{nullflat}), 
the dilatino variation gives the condition, 
\be
\label{codnsusy}
\gamma^+ \epsilon=0. 
\ee
The gravitino variations take the form, 
\begin{subequations}
\begin{align}
\label{grvar}
\kappa \delta \psi_i &  = \partial_i\epsilon -\frac{1}{8} \gamma_i\gamma_\omega (1-\Gamma^4)\epsilon 
-\frac{1}{4} f' \gamma^i \gamma_- \epsilon \\
\kappa \delta \psi_m & = \partial_m \epsilon + \frac{1}{8} \epsilon \,\omega_m -
\frac{1}{8} \gamma_\omega\gamma_m(1-\Gamma^4)\epsilon \\
\kappa \delta \psi_- & = \partial_-\epsilon - \frac{1}{8}\gamma_-\gamma_\omega(1-\Gamma^4) \epsilon \\
\label{grvar+}
\kappa \delta \psi_+ & = \partial_+\epsilon -\frac{1}{8} \gamma_+\gamma_\omega (1-\Gamma^4) \epsilon
 -\frac{1}{4} f' \epsilon + \frac{1}{4} f' \gamma^- \gamma_- \epsilon,
\end{align}
\end{subequations}
where $\omega_m=\partial_m\ln Z, \gamma_\omega=\gamma^m\omega_m$, and
$\Gamma^4=i\Gamma^{0123}$.

It is easy to see that all the conditions are satisfied if and only if
\begin{equation}\label{condsusy}
\begin{aligned}
\Gamma^4\epsilon & = \epsilon, & \qquad
\gamma^+\epsilon & = 0, & \qquad
\epsilon & = Z^{-1/8}e^{f/4} \eta, 
\end{aligned}
\end{equation}
for $\eta$ a  constant spinor.  (Equivalently, $\gamma^+ \eta = 0$
and $i \gamma^{23} \eta = \eta$.)
The first condition in Eq.~(\ref{condsusy})  is the standard one for 
supersymmetry in the presence of a D3 brane and gives rise to $16$ 
supersymmetries. We see that the second condition in Eq.~(\ref{condsusy}) 
is the same as Eq.~(\ref{codnsusy}) and breaks the supersymmetry further 
by half giving rise to a total of  $8$ supersymmetries. 
Thus all backgrounds of the form, Eq.~(\ref{nullflat}) preserve $8$ 
supersymmetries.

\section{Some Comments on the Dual Gauge Theory}

Here we take a few preliminary steps in constructing and analysing the
gauge theory duals to the supergravity backgrounds discussed above.

We would like to claim that the backgrounds discussed in
section~\ref{sec:soln}, Eqs.~\eqref{geom} and~(\ref{conds}), are dual
to an ${\mathcal N}=4$ gauge theory with a gauge coupling
$g_{YM}^2=e^\phi$, in a four dimensional spacetime with metric,
${\tilde g}_{\mu\nu}$.  We now give some supporting evidence for this
claim.

The $AdS_5\times S^5$ background is a special case of Eq.~(\ref{geom}) with 
${\tilde g}_{\mu\nu}=\eta_{\mu\nu}$, and a constant dilaton. Small 
fluctuations around this background give rise to supergravity modes.
These can be mapped to operators using the AdS/CFT dictionary. The family of 
solutions, Eqs.~(\ref{geom}) and~Eq.~(\ref{conds}), include some of these 
supergravity modes as well. 
For example, as was discussed in section 2, solutions where the dilaton 
varies satisfying Eq.~(\ref{conds}), with the metric 
${\tilde g}_{\mu\nu}=\eta_{\mu \nu}$, is a solution of the linearised 
equations obtained from Eqs.~(\ref{geom}) and~(\ref{conds}). 
This mode is dual to the operator ${\rm{Tr}} F^2$ in the Yang -Mills theory. 
Similarly the mode where 
${\tilde g}_{\mu\nu}=\eta_{\mu \nu}+h_{\mu\nu}(x^\mu)$ 
with constant  $\phi$, where $h_{\mu\nu}$ satisfied the linearised Ricci 
flatness condition, ${\tilde R}_{\mu\nu}=0$, is dual to the stress energy 
tensor in the gauge theory, $T_{\mu\nu}$. 
A general linearised solution, $\delta \phi(x^\mu), h_{\mu\nu}(x^\mu)$,
of the type, Eq.~(\ref{geom}), Eq.~(\ref{conds}), then is dual to turning 
on sources in the gauge theory which couple to ${\rm{Tr}} F^2, T_{\mu\nu}$,
\be
\label{sources}
S_{source}=\int d^4x [\delta \phi(x^\mu) {\rm{Tr}} F^2 + h_{\mu\nu}T^{\mu\nu}],
\ee 
which is in agreement with the claim above. 

Most solutions, especially the interesting ones we have found above,
are of course not small fluctuations.  Since the identification of the
dilaton and ${\tilde g}_{\mu\nu}$ with the gauge coupling and metric
of the gauge theory works for the small perturbations, it seems
reasonable to assert that this is true for these solutions as well.

One additional piece of evidence here is to consider a single D3 brane
moving in the background of such a solution. It is easy to see from
the DBI action of this brane that it has a gauge coupling
$e^{\phi/2}$, and metric, ${\tilde g}_{\mu\nu}$. Note that it is only
for D3 branes that the excitations perceive the Einstein metric
${\tilde g}_{\mu\nu}$.

Now actually for small fluctuations, each supergravity mode has two
solutions which fall off differently as $r\rightarrow \infty$. These
correspond to the non-normalisable and normalisable modes
\cite{vijay9805, vijay9808} and determine the source coupling to the
dual operator and the expectation value of the operator in the dual
theory respectively.  It is easy to see that the linearised
perturbations which lie in the class of solutions, Eqs.~(\ref{geom})
and~(\ref{conds}), correspond to only exciting the non-normalisable
modes, as may be seen from the positive power of $r$ in these
perturbations. Thus, in their case the dual gauge theory continues to
be in the ${\mathcal N}=4$ vacuum with the sources mentioned above
turned on.  What happens for solutions which are not small
fluctuations is less clear.  Solutions which approach $AdS_5 \times
S^5$ for early times, when $t \rightarrow -\infty$ or $X^+ \rightarrow
-\infty$ must be dual to the gauge theory starting in the ${\mathcal
N} =4$ vacuum at early times.  The subsequent state of the gauge
theory would then be determined by the sources which are turned on.

Once we have accepted the identification proposed above we can analyse
the gauge theory description in some more detail. The most interesting
question is whether the gauge theory dual to a singular spacetime is
itself singular or not.  In the time dependent solutions we analysed,
of Kasner type, Eq.~(\ref{egconds}), at the cosmological singularity
in the bulk the four dimensional metric ${\tilde g}_{\mu\nu}$ is also
singular. This suggests that the gauge theory living in such a
singular spacetime would itself be singular and ill-behaved.  One case
worth examining separately is when ${\tilde g}_{\mu\nu}$ is
conformally flat.  An example is the solution, Eq.~(\ref{egconds}),
with $p_i={1\over 3}, \alpha={2\over \sqrt{3}}$.  Since the gauge
theory is conformally invariant it might seem at first glance that it
is oblivious of the shrinking conformal factor. This is of course not
true. The conformal anomaly in 4 dimensions \cite{duff20yrs, osborn9307, 
anselmi9601, erdmenger9605, anselmi9708} (see also \eg\ 
\cite{skenderis9806, skenderis9812, balakraus9902, magoo9905} in the 
holographic context of $AdS_5\times S^5$) tells us that 
\be
\label{ca}
T_{\mu}{^\mu} = 
{c\over 16\pi^2}(C_{\alpha\beta\gamma\delta}C^{\alpha\beta\gamma\delta}) 
- {a\over 16\pi^2}(R_{\alpha\beta\gamma\delta}R^{\alpha\beta\gamma\delta} 
- 4R_{\alpha\beta}R^{\alpha\beta} + R^2)\ \propto\ \ 
-R_{\alpha\beta}R^{\alpha\beta} + {1\over 3}R^2\ ,
\ee
where in the last expression we have used $a=c={N^2-1\over 4}$ for the 
$SU(N)$ \Nf\ super Yang Mills theory. Note that in the first expression, 
the first term involves the Weyl tensor and vanishes in a conformally 
flat space-time. Focussing on the \Nf\ theory at hand, the terms in the 
last expression give $T_\mu{^\mu} \propto 1/t^4$ in the example above. 
Thus we see that the stress energy tensor blows up at the singularity in 
this example signalling that the gauge theory is probably ill behaved.

The null solutions are more promising in this respect. These solutions
preserve $8$ supercharges, as we discussed above.  In the conformally
flat cases, Eq.~(\ref{nullflat}), since $R_{++}$ is the only
non-vanishing component of the stress tensor, the conformal anomaly
vanishes.

Consider in particular the solution discussed in
Eq.~(\ref{nullsolntanh}). In this case the solution approaches $AdS_5
\times S^5$ as $X^+ \rightarrow \pm \infty$. Thus the sources in the
gauge theory are turned off at $X^+=\pm \infty$ where it becomes the
${\mathcal N}=4$ Yang Mills theory.  We also learn, as was mentioned
above, that the gauge theory is in the ${\mathcal N}=4$ vacuum as
$X^{+} \rightarrow -\infty$.  
We expect that the deformed gauge theory
inherits the supersymmetries of the bulk since it is basically the
theory on the branes which themselves give rise to the background
and one might hope that this may be useful in drawing conclusions
about the nature of the state at finite $X^+$. At
$X^{+} \rightarrow -\infty$, the number of supersymmetries are
enhanced to the maximal number and the gauge theory dual should be in
the state annihilated by all the supercharges. At finite time one may
naively think that the state continues to be annihilated by eight of
the supercharges. However, all these supersymmetry parameters obey
$\gamma^+ \epsilon = 0$. The corresponding supercharges do not commute
with the hamiltonian and therefore a state which preserves these
supersymmetries at some time alone do not generically preseve this at
later times.  In fact the anticommutator of the corresponding
supercharges is proportional to $P_-$ rather than the hamiltonian.

We have not been able to conclusively establish yet whether the gauge
theory continues to be non-singular as one approaches $X^+=0$ and
leave this for further study \cite{WiP}. If true, we should be able to
answer whether the solution of the form, Eq.~(\ref{nullsolntanh}), is
also valid in the far future, as $X^+ \rightarrow \infty$, or what is
its appropriate continuation in that region.
 
\acknowledgments

It is a pleasure to thank A.~Awad, S. Minwalla, T. Wiseman, X. Wu and C. 
Zhou for discussions. KN thanks the hospitality of the KITP Santa Barbara, 
USA, and the Harvard Theory Group, USA, in the course of this work. SPT 
acknowledges support from the Swarnajayanti Fellowship, DST, Government of 
India. S.R.D. would like to thank Tata Institute of Fundamental Research 
for hospitality during various stages of this work. We especially thank 
the people of India, USA and Canada for generously supporting research in 
string theory.  This work was supported in part by Department of Energy 
contract \#DE-FG01-00ER45832 and the National Science Foundation grant No. 
PHY-0244811.


\end{document}